# Operator Preparation and Characteristic Analysis of Open Quantum Systems Based on the Lyapunov Control Method

Jie Wen, Shuang Cong*
Department of Automation, University of Science and Technology of China,
Hefei, China
*Corresponding Author's E-mail: scong@ustc.edu.cn

**Abstract:** The control laws based on quantum Lyapunov control method are designed to prepare operators for two level open quantum systems in this paper. A novel Lyapunov function $V$ is proposed according to a matrix logarithm function. The higher accuracy and faster convergence of the novel Lyapunov function proposed are analyzed by comparing with the operator distance $V_{dis}$. We adopt three control fields, and design two types of control law forms. The first form is to use one control law to offset the influence of the dissipation, while the other one is to use all three control laws to offset the dissipation. Moreover, the robustness of the system Hamiltonian with uncertainty is further investigated to comprehend the performances of control laws. NOT gates are prepared by the designed control laws for open quantum systems as well as closed quantum systems in numerical experiments to verify the superiority of $V$, in which the performance indexes of distance and fidelity are compared and analyzed.
**Key words**: open quantum systems, operator preparation, Lyapunov function, robustness

## 1. Introduction

In order to realize a quantum computer, one needs to build quantum circuit with different functions of the quantum logic gates to carry around and manipulate the quantum information, so the preparation of quantum gates is a current research focus. A high enough coherence of a quantum system during working time is a necessary condition for an actual quantum system [1-2]. However, the coherence time in physical systems may be limited by the undesired interactions between the qubit and the environment as well as between the qubits itself [3]. These interactions can reduce the gate fidelity or prevent the preparation of quantum gate. Many control methods have been proposed and used in order to prepare quantum gates faster in high fidelity within coherence time. For different types of systems, the usual method of obtaining quantum gates is based on optimal control theory. The optimal control method is applied in a wide range, not only the closed quantum systems which neglect the coupling to the environment [4-7], but also the open quantum systems which interact with the environment [8-14]. Another usual method of obtaining quantum gates is dynamical decoupling [15, 16]. Besides, one can also prepare the high fidelity quantum gates by combining optimal control theory and dynamical decoupling [17]. Although the mentioned methods can prepare high fidelity quantum gate, either optimal control method or dynamical decoupling, the designed controls are not analytic which gives rise to many difficulties in further studies, analyses of system characteristics and actual system realization. Relative to those methods, the control laws designed by Lyapunov method have explicit mathematical expressions and the design process is simpler, so that there are certain advantages in control parameters adjustment and system characteristics analysis. Meanwhile, the Lyapunov method is also appropriate for time-varying systems and nonlinear systems, and the designed control laws can ensure the stability of control systems, thereby this is a simple and effective method to design control laws. For quantum systems, quantum Lyapunov method has been used to the state transfer [18-22], trajectory tracking [23] and so on.

We will employ this method to the operator preparation of two level open quantum systems. The main goal of this paper is to design the control laws which are used to prepare the operators based on the Lyapunov method. In the procedure of the Lyapunov control design, one important thing is to find or construct a suitable Lyapunov function. For the operators preparation, a few functions can be selected as a Lyapunov function in which a nature choice is the distance $V_{dis}$ between the evolution operator $U(t)$ and the target operator $U_f$. We construct a novel Lyapunov function $V$, and it can be seen from this paper that the constructed $V$ has advantages in numerical accuracy and convergence speed relative to the conventional $V_{dis}$. The control laws based on $V$ are suitable for open quantum systems including Markovian quantum systems, Non-Markovian quantum systems, as well as closed quantum systems. Furthermore, we find that the different forms of control laws designed based the same Lyapunov function indicate different performance characteristics in preparing operators. On these foundations, the robustness of system Hamiltonian with uncertainty is investigated, and a way for the control laws to enhance the system robustness is proposed. Besides, it should be noticed that quantum gates are a special class of operators, so the preparation of quantum gates belongs to the preparation of operators, means the preparation of quantum gates is included in this paper.

The remainder of the paper is organized as follows. In Section 2, we introduce the mathematical model of open quantum systems and define an operator purity. In Section 3, we present the constructed Lyapunov function $V$ in detail and analyze the relationship between $V$ and $V_{dis}$, then design the control laws based on $V$ by Lyapunov method to prepare the operators, and we further investigate the robustness of the system Hamiltonian with uncertainty. In Section 4, we select NOT gate as the target operator and carry out numerical simulation experiments under the designed control laws. The conclusions are summarized in Section 5.

## 2. System model description and operator characteristic analysis

In this section, we describe the system model and obtain the operator dynamical equation. Moreover, we also define an index operator purity to judge the system types based on the operator characteristics.

### 2.1 Description of system model

The state density matrix of a two level Markovian quantum system satisfies the following Lindblad equation [24]

$$\dot{\rho}(t) = -i[H(t), \rho(t)] + L(\rho(t)) \qquad (1)$$

where, $H(t) = \frac{1}{2}(f_x(t)\sigma_x + f_y(t)\sigma_y + f_z(t)\sigma_z)$ is the system Hamiltonian, $f_x(t)$, $f_y(t)$ and $f_z(t)$ are the control laws, $\sigma_x = \begin{bmatrix} 0 & 1 \\ 1 & 0 \end{bmatrix}$, $\sigma_y = \begin{bmatrix} 0 & -i \\ i & 0 \end{bmatrix}$ and $\sigma_z = \begin{bmatrix} 1 & 0 \\ 0 & -1 \end{bmatrix}$ are Pauli matrices;

$-i[H(t),\rho(t)] = -i(H(t)\rho(t) - \rho(t)H(t))$ is the unitary part of evolution, and the dissipative part $L(\rho(t))$ of the evolution is

$$L(\rho(t)) = \sum_{\alpha\beta} \gamma_{\alpha\beta} \left( F_\alpha \rho(t) F_\beta^\dagger - \frac{1}{2}\left(F_\beta^\dagger F_\alpha \rho(t) + \rho(t) F_\beta^\dagger F_\alpha\right)\right) \tag{2}$$

where, $F_\alpha \in \frac{1}{\sqrt{2}}\{\sigma_x, \sigma_y, \sigma_z\}$ [25, 26]; $\gamma_{\alpha\beta}$ indicates the coupling strengths of the system with the environment, and forms a Hermitian and positive semi-definite matrix

$$\Gamma = \begin{bmatrix} \gamma_{xx} & \gamma_{xy} & \gamma_{xz} \\ \gamma_{yx} & \gamma_{yy} & \gamma_{yz} \\ \gamma_{zx} & \gamma_{zy} & \gamma_{zz} \end{bmatrix} \tag{3}$$

which is known as the GKS (Gorini, Kossakowski and Sudarshan) matrix [27]. When the entries of $\Gamma$ are time-dependent, the system becomes a Non-Markovian system while the system becomes a closed quantum system when $\Gamma = \mathbf{0}$.

Generally, there are two types of the most widely used Markovian quantum system models: phase damping (short for PD) model and amplitude damping (short for AD) model, respectively. The PD system describes a decoherence process which can be caused by random phase shifts of the system due to its interaction with the environment [28]. The dissipative part of PD system in (1) is [29]:

$$L_{PD}(\rho) = -\frac{\gamma}{4}[\sigma_z, [\sigma_z, \rho]] \tag{4}$$

According to (3) and (4), the corresponding GKS matrix is $\Gamma_{PD} = \gamma \cdot diag(0,0,1)$. The AD system describes the process in which the initial state $|1\rangle$ will gradually decay to the ground state $|0\rangle$ [28]. The dissipative part of AD system in (1) is [29]:

$$L_{AD}(\rho) = \frac{\gamma}{2}\left(\sigma_-\rho\sigma_+ - \frac{1}{2}(\sigma_+\sigma_-\rho + \rho\sigma_+\sigma_-)\right) \tag{5}$$

where $\sigma_\pm = \sigma_x \pm i\sigma_y$. According to (3) and (5), the corresponding GKS matrix is $\Gamma_{AD} = \gamma \cdot \begin{bmatrix} 1 & i & 0 \\ -i & 1 & 0 \\ 0 & 0 & 0 \end{bmatrix}$.

Moreover, the dissipative part of a specific Non-Markovian quantum systems can be [30, 31]

$$L_{NM}(\rho) = \sum_{k=0}^{1} d_k(t)\left(2\sigma_k \rho \sigma_k^+ - \sigma_k^+ \sigma_k \rho - \rho \sigma_k^+ \sigma_k\right) \tag{6}$$

where $\sigma_0 = \frac{\sigma_+}{2}$, $\sigma_1 = \frac{\sigma_-}{2}$; attenuation coefficients $d_k(t)$ are

$$d_0(t) = d_1(t) = d(t) = \alpha^2 kT \frac{\beta^2}{1+\beta^2}\left(1 - e^{-\beta\omega_0 t}\left(\cos(\omega_0 t) - \frac{1}{r}\sin(\omega_0 t)\right)\right) \tag{7}$$

where, $\alpha^2 = 0.01$ is the coupling strength between the system and the environment; $\beta = \omega_z/\omega_0$ denotes the ratio of bath cutoff frequency $\omega_z$ to system oscillator frequency $\omega_0$; $kT = 300$. When $d_0(t)$ and $d_1(t)$ are all constant more than zero, namely the system decays at a constant speed, the

Non-Markovian quantum systems becomes a Markovian quantum system. Specially, the Non-Markovian quantum systems becomes AD system when $d_0(t) = d_1(t) = \gamma/2$.

In order to prepare operators, one needs to obtain the operator dynamics first. Due to the dissipative part $L(\rho(t))$, the state density dynamics (1) is a nonlinear equation, it is difficult to derivate and obtain the operator dynamics. The two level quantum systems can also be described by state vector by which the state dynamics becomes a linear equation. In this case, it is easier to research the system characteristics and make mathematical derivation. Thus we will convert the density matrix into the state vector and obtain the operator dynamics by the state vector dynamics instead of state density dynamics.

Let $\rho = \frac{1}{2}I + \frac{1}{2}(r_x\sigma_x + r_y\sigma_y + r_z\sigma_z)$, and place $\rho$ into (1), then the state vector $r = (r_x, r_y, r_z)^T$ satisfies the equation

$$\dot{r}(t) = (A(t) + B)r(t) \tag{8}$$

where,

$$A(t) = \begin{bmatrix} 0 & -f_z(t) & f_y(t) \\ f_z(t) & 0 & -f_x(t) \\ -f_y(t) & f_x(t) & 0 \end{bmatrix} = f_x(t)A_x + f_y(t)A_y + f_z(t)A_z \tag{9}$$

which corresponds to the unitary part $-i[H(t), \rho]$, and $A_x = \begin{bmatrix} 0 & 0 & 0 \\ 0 & 0 & -1 \\ 0 & 1 & 0 \end{bmatrix}$, $A_y = \begin{bmatrix} 0 & 0 & 1 \\ 0 & 0 & 0 \\ -1 & 0 & 0 \end{bmatrix}$,

$A_z = \begin{bmatrix} 0 & -1 & 0 \\ 1 & 0 & 0 \\ 0 & 0 & 0 \end{bmatrix}$.

$$B = \frac{\Gamma + \Gamma^T}{2} - tr(\Gamma)I = \frac{1}{2}\begin{bmatrix} -2(\gamma_{yy} + \gamma_{zz}) & \gamma_{xy} + \gamma_{yx} & \gamma_{xz} + \gamma_{zx} \\ \gamma_{yx} + \gamma_{xy} & -2(\gamma_{xx} + \gamma_{zz}) & \gamma_{yz} + \gamma_{zy} \\ \gamma_{zx} + \gamma_{xz} & \gamma_{zy} + \gamma_{yz} & -2(\gamma_{xx} + \gamma_{yy}) \end{bmatrix} \tag{10}$$

which corresponds to the dissipative part $L(\rho(t))$. $B$ is symmetric with the eigenvalues as $\lambda_1 = -(\mu_2 + \mu_3)$, $\lambda_2 = -(\mu_1 + \mu_3)$, and $\lambda_3 = -(\mu_1 + \mu_2)$ which satisfy $\lambda_1 \geq \lambda_2 \geq \lambda_3$. $\mu_1$, $\mu_2$ and $\mu_3$ are the eigenvalues of $\Gamma$ and satisfy $\mu_1 \geq \mu_2 \geq \mu_3$.

Notice that Equation (8) is a linear equation, and

$$r(t) = U(t)r(0) \tag{11}$$

in which $U(t)$ is an evolution operator. Substituting (11) into (8), one can obtain that the operator $U(t)$ satisfies the following equation:

$$\dot{U}(t) = (A(t) + B)U(t) \tag{12}$$

Now the task of the quantum operator preparation becomes to design the control laws in $A(t)$ of (9) so that $U(t)$ which satisfies the dynamics (12) can as close as possible to the target operator $U_f$ in the shortest possible time.

Moreover, the unitary is a necessary and only condition for operator as a quantum gate. Any unitary operator specifies a valid quantum gate [28]. Thereby we focus on the unitary operator though the operator may be non-unitary in evolution of open quantum systems. When the unitary target operator acting on density matrix $\rho(t)$ is $U_{2\times 2} = \begin{bmatrix} u_1 & u_2 \\ u_3 & u_4 \end{bmatrix}$, then the corresponding operator acting on state vector $r = (r_x, r_y, r_z)^T$ is

$$U_{3\times 3} = \frac{1}{2} \begin{bmatrix} u_1 u_4^* + u_1^* u_4 + u_2 u_3^* + u_2^* u_3 & \left(u_1^* u_4 - u_1 u_4^* + u_2 u_3^* - u_2^* u_3\right)i & 2\left(u_1 u_3^* + u_1^* u_3\right) \\ \left(u_1^* u_4 - u_1 u_4^* + u_2 u_3^* - u_2^* u_3\right)i & u_1^* u_4 + u_1 u_4^* - u_2 u_3^* - u_2^* u_3 & 2\left(u_1 u_3^* - u_1^* u_3\right)i \\ 2\left(u_1 u_2^* + u_1^* u_2\right) & 2\left(u_1^* u_2 - u_1 u_2^*\right)i & 2\left(u_1 u_1^* - u_2 u_2^*\right) \end{bmatrix} \quad (13)$$

where, $u^*$ represents the conjugate of $u$. Taking NOT gate $U_{Not} = \begin{bmatrix} 0 & 1 \\ 1 & 0 \end{bmatrix}$ for example, $U_{Not}$ corresponds to $U_{f-Not} = diag(1, -1, -1)$.

It should be noted that the transformation from $U_{2\times 2}$ to $U_{3\times 3}$ is not a strict one-to-one correspondence. For instance, $U_{2\times 2} = I_{2\times 2}$ and $-I_{2\times 2}$ all correspond to $U_{3\times 3} = I_{3\times 3}$ while $-I_{2\times 2} = e^{i\pi} \cdot I_{2\times 2}$, i.e. $-I_{2\times 2}$ and $I_{2\times 2}$ differ a global phase. Because the global phase cannot be measured, the transformation in (13) can be regarded as one to one.

## 2.2 Operator purity

The dynamics difference between closed quantum systems and open quantum systems is that the dynamics of open quantum systems contain both unitary part and dissipative part, while the closed quantum systems contain only unitary part. Usually, whether a system is a closed quantum system or not can be determined based on the change rules of state purity. However, the operator is studied in this paper, so an operator purity should be defined based on the similar idea of the state purity, according to which the system types can be determined. The state purity $P_{state}$ of two level quantum systems is the distance between a state $r(t) = (r_x(t), r_y(t), r_z(t))^T$ and the state $r_0 = (0,0,0)^T$ which is the original point of the Bloch sphere, i.e. $P_{state} = \sqrt{r_x^2 + r_y^2 + r_z^2}$. The operator $U_0 = \mathbf{0}_{3\times 3}$ can transfer a $r(t)$ to $r_0$, so $U_0$ can be regard as the similar role as $r_0$ in state purity. Correspondingly, we can define the operator purity using the quadratic sum of the differences of each element of an operator $U(t)$ from the corresponding element in $U_0$ as

$$P = \sum_{i=1}^{9} (u_i(t) - 0_i)^2 = \sum_{i=1}^{9} (u_i(t))^2 \quad (14)$$

in which $u_i(t)$ is the element of $U(t)$.

The operator purities of different types of systems in free evolution are shown in Fig. 1, in which the operator purity of a closed quantum system has the same characteristic as the state purity has, which remains unchanged and is consistent. The operator purities of the PD and AD Markovian quantum systems decrease monotonically, in which the decrease of an AD system is more greatly. The operator purity of Non-Markovian quantum system decreases oscillatorily. These different characteristics of open quantum systems are caused by the existence of dissipation, and are the same as the change rules of the state purities. Thus $P$ is used as an index in this paper to determine the system types based on the operator characteristics.

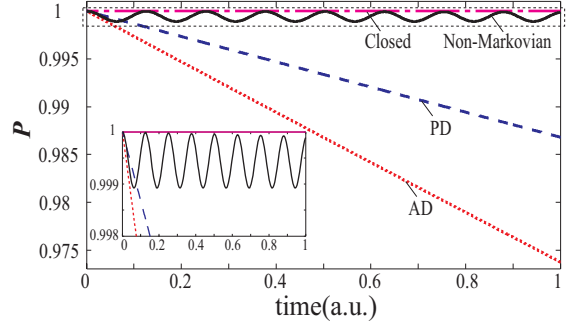

Fig. 1. Operator purities are as the functions of different types of systems in free evolution, in which the dot chain line, the dot dashed line and the long dashed line represent the purities of closed quantum systems, PD system and AD system, respectively. The solid line represents the purity of Non-Markovian quantum system which is also shown in partial enlarged view.

## 3. Design of the control laws and robustness analysis

In this section, we will construct a novel Lyapunov function $V$ and analyze the relationship between the constructed $V$ and the $V_{dis}$ which describes the operator distance, and then design the control laws of preparing operators based on $V$ by Lyapunov method. Moreover, we also analyze the robustness of the system Hamiltonian with uncertainty in this section.

### 3.1 Construction and characteristics analysis of the novel Lyapunov function

The procedure of designing control laws by the Lyapunov method is: to find or construct a suitable Lyapunov function $V$ which satisfies $V \geq 0$ first, and then design control laws so that $V$ decreases monotonically, that is, $\dot{V} \leq 0$. Therefore, the first step is to select the Lyapunov function. For the operator preparation, many functions can be selected as the Lyapunov function in which a relatively natural choice that can be considered intuitively is the distance $V_{dis}(t)$ between the evolution operator $U(t)$ and the target operator $U_f$, i.e.

$$V_{dis}(t) = \left\| U(t) - U_f \right\|^2 \tag{15}$$

In order to obtain higher preparation accuracy and shorter preparation time, Pierre de Fouquieres implemented quantum gates by optimal control with doubly exponential convergence for closed quantum systems, in which the performance function is a matrix logarithm function $\log\left(U_f^\dagger U(t)\right)$ [12]. For the same purpose, we use Lyapunov control method instead of optimal control method and construct a novel Lyapunov function $V$ which is from the same matrix logarithm function for open quantum systems in this subsections.

For the sake of writing concisely, denote $W(t) = U_f^\dagger U(t)$. When the spectral radius of $W(t) - I$ is less than 1, i.e. $\rho(W(t) - I) < 1$, the Mercator's series of $\log(W(t))$ are [32]:

$$\log(W(t)) = (W(t) - I) - \frac{1}{2}(W(t) - I)^2 + \frac{1}{3}(W(t) - I)^3 - \frac{1}{4}(W(t) - I)^4 + \ldots \tag{16}$$

We take the norm square of the first two items of the Mercator's series in (16) as the Lyapunov function $V$, i.e.

$$V(t) = \|L(t)\|^2 = tr\left(L^\dagger(t)L(t)\right) \tag{17}$$

where, $L(t) = (W(t)-I) - \frac{1}{2}(W(t)-I)^2$.

One can see from (17) that $V(t) \geq 0$ at any time $t$ and equal to zero when $U(t) = U_f$, i.e. $W(t) = I$ which means $V(t)$ in (17) satisfies the requirements of Lyapunov function. Besides, $V_{dis}$ can be rewritten as

$$V_{dis} = 2N - 2\operatorname{Re} tr(W(t)) = \|W(t)-I\|^2 \tag{18}$$

where, $N$ is the system dimension.

One can see easily by comparing (18) with (16) that conventional $V_{dis}$ is the norm square of the first item of the Mercator's series of $\log(W(t))$. According to (17) and (18) as well as their first order time derivatives, the relationships between $V$ and $V_{dis}$ in numerical value and convergence speed are:

1) The relationship between $V$ and $V_{dis}$ in numerical value is

$$V = \frac{13}{4}V_{dis} + \frac{3}{2}\left(\operatorname{Re} tr(W^2) - \operatorname{Re} tr(W)\right) \tag{19}$$

Thus $V = f_1(tr(W), tr(W^2))$ while $V_{dis} = f_2(tr(W))$, i.e. $V$ is a function of two variables $tr(W)$ and $tr(W^2)$ while $V_{dis}$ is a function of one variables $tr(W)$. Specially, when $U(t) \to U_f$, namely $W \to I$, let $W = I + \Delta I$ in which $\Delta I$ is diagonal matrix with diagonal element $e$, $V_{dis}$ and $V$ are

$$\begin{aligned} V_{dis} &= -2N \cdot e \\ V &= \frac{5}{2}V_{dis} - \frac{3}{4}V_{dis} \cdot e = -5N \cdot e + \frac{3}{2}N \cdot e^2 \end{aligned} \tag{20}$$

from which one can see that $V_{dis}$ is only relevant to $e$, while $V$ is relevant to $e$ and $e^2$ so that $V$ has higher accuracy in numerical value.

2) The relationship between $V$ and $V_{dis}$ in convergence speed is

$$\dot{V} = 4\dot{V}_{dis} + 3\operatorname{Re} tr(W \cdot \dot{W}) \tag{21}$$

Likewise, when $U(t) \to U_f$, namely $W \to I$, let $W = I + \Delta I$, $\dot{V}_{dis}$ and $\dot{V}$ are

$$\begin{aligned} \dot{V}_{dis} &= -2\operatorname{Re} tr(\dot{W}) \\ \dot{V} &= \frac{5}{2}\dot{V}_{dis} - \frac{3}{2}\dot{V}_{dis} \cdot e \end{aligned} \tag{22}$$

According to (20) and (22), the convergence speed of the first item in $V$ is identical with $V_{dis}$, while the convergence speed of the second item is twice as much as $V_{dis}$, thereby the total convergence speed of $V$ is faster than that of $V_{dis}$.

One can see from the comparison of $V$ and $V_{dis}$ that $V$ has higher numerical accuracy and faster convergence speed relative to $V_{dis}$, thus we use $V$ as the Lyapunov function in this paper, and design the control laws of preparing operators based on $V$ in the next subsection.

**3.2 Design of Control laws**

After the construction of the Lyapunov function $V(t)$, we begin to design control laws so as to let

$V(t)$ decrease monotonically, i.e. $\dot{V}(t) \leq 0$. The first order time derivative of $V(t)$ is

$$\dot{V}(t) = tr\left(\dot{L}^\dagger(t)L(t) + L^\dagger(t)\dot{L}(t)\right)$$
$$= tr\left(\left(-\frac{1}{2}\dot{U}^\dagger U_f W^\dagger - \frac{1}{2}W^\dagger \dot{U}^\dagger U_f + 2\dot{U}^\dagger U_f\right)L + L^\dagger\left(-\frac{1}{2}U_f^\dagger \dot{U} W - \frac{1}{2}W U_f^\dagger \dot{U} + 2U_f^\dagger \dot{U}\right)\right) \quad (23)$$
$$= f_x(t)S(A_x) + f_y(t)S(A_y) + f_z(t)S(A_z) + S(B)$$

in which $S(X) = 2\operatorname{Re} tr\left(\left(-\frac{1}{2}U^\dagger X^\dagger U_f W^\dagger - \frac{1}{2}W^\dagger U^\dagger X^\dagger U_f + 2U^\dagger X^\dagger U_f\right)L\right)$.

In order to ensure $\dot{V}(t) \leq 0$, the control laws can be designed as

$$\begin{aligned} f_x(t) &= -\frac{S(B)}{S(A_x)} \\ f_y(t) &= -k_y S(A_y) \\ f_z(t) &= -k_z S(A_z) \end{aligned} \quad (24)$$

where, $k_y, k_z \geq 0$. $f_x(t)$ is used to offset the dissipation caused by $B$ in (12) while $f_y(t)$ and $f_z(t)$ are used to prepare operators.

It can be seen from (24) that the mathematical expressions of $f_y(t)$ and $f_z(t)$ are similar, so they have similar properties. The $f_x(t)$ contains denominator $S(A_x)$. In the case when $S(A_x)$ is equal or close to zero, the amplitude of $f_x(t)$ will tends to infinite. Hence one needs to limit the amplitude of $f_x(t)$ appropriately in the actual experimental use. In order to do so, when the amplitude is more than a certain value, one needs to use the value of the previous time to replace the value of current time. Placing the control laws (24) into (23), one can get

$$\dot{V}(t) = -k_y \left(S(A_y)\right)^2 - k_z \left(S(A_z)\right)^2 \leq 0 \quad (25)$$

One can verify the control laws (24) may ensure $\dot{V}(t) \leq 0$, too.

Obviously, the control laws which can make $\dot{V}(t) \leq 0$ are not unique, in which (24) is the relatively simple one. Another kind of control laws that can be considered and make $\dot{V}(t) \leq 0$ is to add control components which are used to offset $B$ to the three control laws at the same time, i.e.

$$\begin{aligned} f_{nx}(t) &= -k_{nx}S(A_x) - h_{nx} \cdot \frac{S(B)}{S(A_x)} \\ f_{ny}(t) &= -k_{ny}S(A_y) - h_{ny} \cdot \frac{S(B)}{S(A_y)} \\ f_{nz}(t) &= -k_{nz}S(A_z) - h_{nz} \cdot \frac{S(B)}{S(A_z)} \end{aligned} \quad (26)$$

where, $k_{nx}, k_{ny}, k_{nz} \geq 0$ and $h_{nx} + h_{ny} + h_{nz} = 1$. Here $-k_{nj}S(A_j)$ and $-h_{nj} \cdot \frac{S(B)}{S(A_j)}, j = x, y, z$ are denoted as excluded denominator part and included denominator part of the control laws. Place (26) into (23), one can get that

$$\dot{V}(t) = -k_{nx}\left(S(A_x)\right)^2 - k_{ny}\left(S(A_y)\right)^2 - k_{nz}\left(S(A_z)\right)^2 \leq 0 \quad (27)$$

which means the control laws (26) can also ensure $\dot{V}(t) \leq 0$.

The weights of the three control laws in (26) to offset $B$ can be adjusted by choosing different parameters $h_{nx}$, $h_{ny}$ and $h_{nz}$. In the three control law expression forms in (26), the action of the included denominator part is to offset dissipation while the action of excluded denominator parts is to prepare operator. When $h_{nx}=1$, $k_{nx}=h_{ny}=h_{nz}=0$, then (26) is the same as (24) which indicates that (24) is a special case of (26) and is less flexibility than (26). However, the control laws (26) contain more included denominator parts which may appear the case in which the denominators are equal or close to zero in the evolution, and the changes of the control amplitude in (26) are more drastic. Although (24) and (26) are all designed based on $V$, their performances and characteristics are different in preparing operators due to the different mathematical expressions, which will be explained and verified by experiments in subsection 4.2.

### 3.3 Robustness analysis

The actual system is usually affected by perturbations from the environment or other sources, so it is a reasonable requirement that the designed control laws should be robust to resist the variation of parameters and the effect of perturbations so that the robustness is an important index to reflect the control law characteristics. We will investigate the robustness of system Hamiltonian with uncertainty in this subsection. The uncertainties can be taken into account by adding a perturbation $\lambda\sigma$ to the system Hamiltonian $H(t)$, i.e. $H(t)$ becomes

$$\tilde{H}(t)=H(t)+\lambda\sigma \tag{28}$$

where $\lambda\sigma=\frac{1}{2}\sum_{k=i,x,y,z}\lambda_k\sigma_k$ with $\lambda_k$ a real number and $\sigma_k\,(k=i,x,y,z)$ the unit matrix and Pauli matrix, respectively. The effects of different $\sigma_k$ on the system are as follow:

1) When $\lambda\sigma=\lambda_i I$, $H(t)$ becomes

$$\tilde{H}_i(t)=H(t)+\frac{1}{2}\lambda_i I \tag{29}$$

one can get easily that $[H(t),\rho(t)]=[\tilde{H}_i(t),\rho(t)]$ so that $A(t)$ in (12) remains unchanged, namely $\frac{1}{2}\lambda_i I$ doesn't affect system, so the robustness is strongest.

2) When $\lambda\sigma=\lambda_x\sigma_x$, $H(t)$ becomes

$$\tilde{H}_x(t)=\frac{1}{2}\left((f_x(t)+\lambda_x)\sigma_x+f_y(t)\sigma_y+f_z(t)\sigma_z\right) \tag{30}$$

from which one can see that $f_x(t)$ becomes $\tilde{f}_x(t)=f_x(t)+\lambda_x$, namely the perturbation lead to a deviation $\lambda_x$ between the actual control law $\tilde{f}_x(t)$ and the designed control law $f_x(t)$ while $f_y(t)$ and $f_z(t)$ are consistent with the theoretical values. At this point, $A(t)$ becomes

$$A_{\lambda x}(t)=\begin{bmatrix} 0 & -f_z(t) & f_y(t) \\ f_z(t) & 0 & -f_x(t)-\lambda_x \\ -f_y(t) & f_x(t)+\lambda_x & 0 \end{bmatrix} \tag{31}$$

3) When $\lambda\sigma$ is $\lambda_y\sigma_y$ and $\lambda_z\sigma_z$, respectively, $H(t)$ becomes

$$\tilde{H}_y(t)=\frac{1}{2}\left(f_x(t)\sigma_x+\left(f_y(t)+\lambda_y\right)\sigma_y+f_z(t)\sigma_z\right),\tilde{H}_z(t)=\frac{1}{2}\left(f_x(t)\sigma_x+f_y(t)\sigma_y+\left(f_z(t)+\lambda_z\right)\sigma_z\right) \quad (32)$$

At this point, the perturbation makes the control laws $f_y(t)$ and $f_z(t)$ become $\tilde{f}_y(t)=f_y(t)+\lambda_y$ and $\tilde{f}_z(t)=f_z(t)+\lambda_z$, respectively, while $f_x(t)$ is consistent with the theoretical values. Correspondingly, $A(t)$ become

$$A_{\lambda y}(t)=\begin{bmatrix} 0 & -f_z(t) & f_y(t)+\lambda_y \\ f_z(t) & 0 & -f_x(t) \\ -f_y(t)-\lambda_y & f_x(t) & 0 \end{bmatrix}, \quad A_{\lambda z}(t)=\begin{bmatrix} 0 & -f_z(t)-\lambda_z & f_y(t) \\ f_z(t)+\lambda_z & 0 & -f_x(t) \\ -f_y(t) & f_x(t) & 0 \end{bmatrix} \quad (33)$$

respectively.

According to (30) and (32), one can see that the uncertainty of system Hamiltonian leads to the following results: there are deviations in numerical values between the actual control laws acting on system and the designed control laws. The effects of the deviations on the control laws will be studied in subsection 4.4.

## 4. Numerical simulations and result analysis

In order to research the properties of the control laws designed based on the proposed Lyapunov function $V$ more fully for different types of systems, the control laws will be used to prepare NOT gates for all types of quantum systems by numerical simulations in this section.

### 4.1 Performance indexes in experiments

The system models used in numerical simulation experiments are the Markovian quantum systems (4) and (5) as well as the Non-Markovian quantum system (6). In order to better compare the characteristics, two performance indexes are introduced to describe the preparation accuracy of the operators. One is the Distance $D$ and the other is Fidelity $F$. They are defined as follows.

1) **Distance $D$:** The distance $D$ between the evolution operator $U(t)$ and the target operator $U_f$ is defined as

$$D=V_{dis}=\|U(t)-U_f\|^2=tr\left((U(t)-U_f)^\dagger\cdot(U(t)-U_f)\right) \quad (34)$$

Let $G(t)=(U(t)-U_f)^\dagger$, then $D=tr(G(t)\cdot G^\dagger(t))$. If $tr(G\cdot G^\dagger)=0$ with $G\in C^{n\times n}$, then $G=\mathbf{0}$ [33]. As a consequence, when $D=0$, one can get $G(t)=\mathbf{0}$, i.e. $U(t)-U_f=\mathbf{0}$ so that $U(t)=U_f$ which means $D$ can be as the index to describe whether $U(t)$ reaches $U_f$. Considering fault-tolerant quantum computation, the necessary condition of a quantum operator being valid is [34]:

$$D<10^{-4} \quad (35)$$

Therefore, (35) is as the criterion of valid operator in this paper.

2) **Fidelity $F$:** The cases in which the operator is non-unitary in the evolution exist for open quantum systems, so the fidelity $F$ which describes the similarity of two operators is defined as [35]

$$F=\frac{tr\left(U(t)U^\dagger(t)\right)+\left|tr\left(U_f^\dagger U(t)\right)\right|^2}{N(N+1)} \quad (36)$$

where $N$ is the system dimensions. The more similar the two operators are, the higher the fidelity is, and otherwise, the lower the fidelity is.

In the numerical simulations, $D$ and $F$ are both used to describe the operator preparation accuracy. We will do the following three experiments in the following subsections:

1) Preparation of the NOT gate for PD system and AD system by the control laws (24) and (26), respectively, to verify that (24) and (26) are all effective for Markovian quantum systems, and analyze their characteristics;

2) Preparation of the NOT gate by (24) for Non-Markovian quantum systems and closed quantum systems to verify that (24) is also effective for both Non-Markovian quantum systems and closed quantum systems;

3) Preparation of the NOT gate by (24) for PD system and AD system when the system Hamiltonian contains uncertainty to investigate the robustness.

**4.2 NOT gate preparation and control performance analysis for Markovian quantum systems**

In this subsection, we will prepare the NOT gates for PD system and AD system as shown in (4) and (5) by control laws (24) and (26), and analyze the control performances and characteristics through the experiments results. In order to analyses characteristics clearly, in the simulations we define two stages: the stage in which $D$ has not reached minimum value is denoted as operator preparation stage, and the stage in which $D$ has reached minimum value and kept below $10^{-4}$ is denoted as operator preservation stage. The $D$ and $F$ are as the function of time in preparing the NOT gates by (24) and (26) for the PD and AD systems are shown in Fig. 2, respectively. The minimum values of $D$ and maximum values of $F$ in Fig. 2 as well as the parameters in (24) and (26) are listed in Table 1.

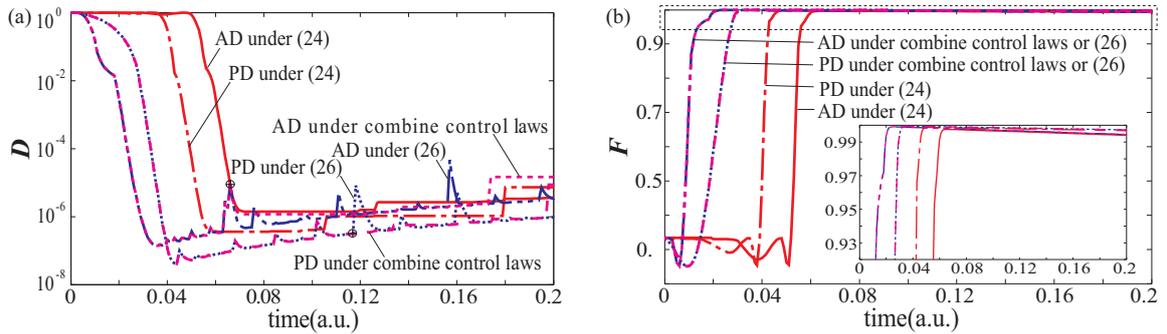

Fig. 2. $D$ and $F$ are as the functions of time in preparing the NOT gates under the controls of (24) and (26) for the PD and AD systems, in which (a) $D$ under (24) and (26) for PD system and AD system, the dot chain line, the dot dashed line and the long dashed line represent the $D$ of preparing NOT gates for the PD system under the control of (24), (26), and combine control (which will be explained in the analysis), respectively. The solid line, two-dot chain line and short dashed line represent the $D$ of preparing NOT gates for the AD system under the control of (24) and (26), respectively; (b) $F$ under (24) and (26) for the PD and AD systems, in which the implications of different types of lines for $F$ are consistent with Fig. (a). The results can be seen from the partial enlarged view more clearly.

From Fig. 2 one can see that 1) the performances of $D$ and $F$ of both AD and PD control systems under the control (26) are superior than those under the control (24); 2) the times of reaching the minimum $D$ and maximum $F$ are less under the control (26); 3) either PD system or AD system under the control (26) has less minimum $D$ and more maximum $F$ than those under the control (24) as shown in Table 1, which means higher accuracy and faster time can be obtained in preparing NOT gates by (26) for both PD and AD systems than by (24). The reasons are: 1) $f_{nx}$ in (26) is used to offset dissipation and prepare operator at the same time, and has more comprehensive action relative to $f_x$ in (24); 2) (24) is a special case of (26) as mentioned in subsection 3.2, more parameters provide more flexibility. The results in Fig. 2 also indicate that the control performances are not only connected with the Lyapunov function, but also the design idea of control laws.

Table 1 Minimum values of $D$ and maximum values of $F$ in Fig. 2 and the parameters in (24) and (26)

| No. | System | Control laws | $f_x(0)$ ($f_{nx}(0)$) | $f_y(0)$ ($f_{ny}(0)$) | $f_z(0)$ ($f_{nz}(0)$) | $k_{nx}$ | $k_y$ ($k_{ny}$) | $k_z$ ($k_{nz}$) | $h_{nx}$ | $h_{ny}$ | Minimum $D$ | Maximum $F$ |
|---|---|---|---|---|---|---|---|---|---|---|---|---|
| 1 | PD | (24) | 99 | 20 | 30 | | 400 | 400 | | | $3.615\times10^{-7}$ | 0.99877 |
| 2 | PD | (26) | 99 | 20 | 30 | 398 | 233 | 59 | 0.21 | 0 | $4.080\times10^{-8}$ | 0.99949 |
| 3 | AD | (24) | 10.28 | 10.73 | 40 | | 400 | 400 | | | $1.390\times10^{-6}$ | 0.99760 |
| 4 | AD | (26) | 10.28 | 10.73 | 40 | 61 | 116 | 397 | 0.35 | 0.31 | $1.357\times10^{-7}$ | 0.99914 |

By means of observing Fig. 2 carefully one can find that the curves of $D$ have many peaks under (26), while there is no any peak under (24) in operator preservation stage. The values of $D$ become bigger and bigger in operator preservation stage. Many peaks whose values are more than $10^{-4}$ appear, which means the performance does not achieve the expected value under the control (26), while not so does under the control (24). A natural and suitable control strategy is to combine superiority of (24) and (26): (26) is used in operator preparation stage while (24) is used in operator preservation stage. The switch point of (26) and (24) is chosen as: the nearest point of indexes when the two control laws act alone, as '⊕' shown in Fig. 2(a). Here the control laws of the strategy are denoted as combine control laws. The results of preparing NOT gates for PD system and AD system by combine control laws are also given in Fig. 2, from which one can see that combine control laws can prepare a both higher accuracy and faster Not gates, meanwhile there are no any peaks in operator preservation stage.

Notice that the above results are related to the fact that $B$ is small in experiments. If $B$ is very big, the values of included denominator parts of control laws will be large in operator prepare stage. Then, which control law should be used in operator preparation and preservation stages is needed to analyze based on the actual case. The big and small of $B$ can be determined according to the proportion of $-S(B)/S(A_k)$ relative to $-S(A_k)$. When the proportion is big, then $B$ is big, otherwise, $B$ is small.

### 4.3 NOT gate preparation and control performance analysis

The two systems studied in subsections 4.2 are all Markovian quantum systems. In order to comprehend the performances of control laws designed based on $V$, we will use the control laws (24) to prepare NOT gates for a Non-Markovian quantum system (6) and a closed quantum system in this subsection.

In the experiments, the parameters $\beta$ and $\omega_0$ in (7) are taken as $\beta = 0.00168$, $\omega_0 = 50$, and $\gamma = 0.1$, respectively. In this setting, the value of attenuation coefficients $d(t)$ in (7) for Non-Markovian systems are closed to the value of $\gamma$ in (4) and (5) for PD system and AD system. $d(t)$ decrease gradually with the time goes which conforms to the characteristics of

Non-Markovian systems. The curves of $D$ and $F$ in preparing the NOT gates under the control (24) for different types of systems are shown in Fig. 3. The minimum values of $D$ and maximum values of $F$ in Fig. 3 as well as the parameters in (24) are listed in Table 2. One can see that from Fig. 3 and Table 2 that minimum values of $D$ are all below $10^{-4}$ and the maximum values of $F$ are all more than 0.997 and close to 1, which means (24) can be applied to all systems including closed quantum systems, Markovian quantum systems and Non-Markovian quantum systems, and preserve the prepared operators to be valid for a longer time.

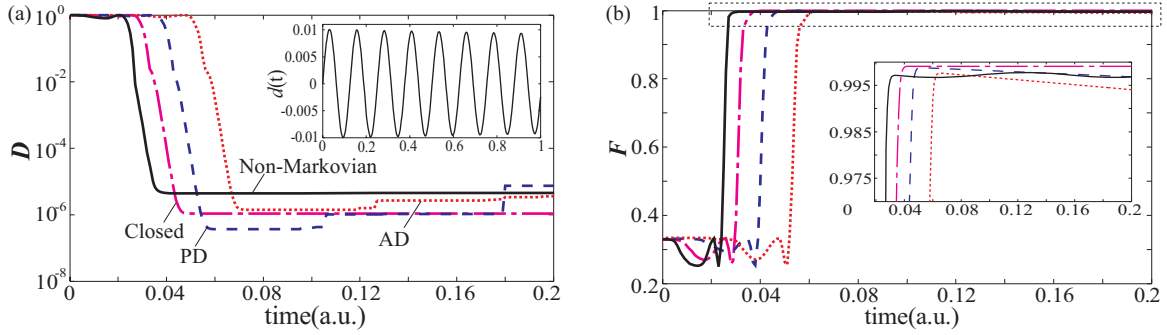

Fig. 3. Curves of $D$ and $F$ in preparing the NOT gates by the control laws (24) for the closed, PD, AD, and non-Markovian systems, in which (a) $D$ under the control (24) for different types of systems; the dot chain line, long dashed line, dot dashed line and solid line represent the $F$ of preparing NOT gates by (24) for closed quantum systems, PD system, AD system and Non-Markovian quantum system, respectively. The curve of attenuation coefficients $d(t)$ is shown in the box; (b) $F$ under (24) for different types of systems; the implications of different types of lines for $F$ are consistent with Fig. (a). The results can be seen from the partial enlarged view more clearly.

Table 2. Minimum values of $D$ and maximum values of $F$ in Fig. 3 and the parameters in (24)

| No. | System | Control laws | $f_x(0)$ | $f_y(0)$ | $f_z(0)$ | $k_y(k_{dy})$ | $k_z(k_{dz})$ | Minimum $D$ | Maximum $F$ |
|---|---|---|---|---|---|---|---|---|---|
| 1 | Closed | (24) | 165.8 | 18.7 | 30 | 400 | 400 | $1.067\times10^{-6}$ | 0.99908 |
| 2 | PD | (24) | 99 | 20 | 30 | 400 | 400 | $3.615\times10^{-7}$ | 0.99877 |
| 3 | AD | (24) | 10.28 | 10.73 | 40 | 400 | 400 | $1.390\times10^{-6}$ | 0.99760 |
| 4 | Non-Markovian | (24) | 135.8 | 3.9 | 136.9 | 400 | 400 | $4.416\times10^{-6}$ | 0.99775 |

In Table 2, the minimum $D$ of PD system is less than that of closed quantum system which means Not gate for PD system has higher accuracy than that for closed quantum system in index $D$. However, the dynamics of PD system contains dissipative part, thus the dissipative part can play positive roles in some cases and is not always adverse.

the difference of the AD and Non-Markovian quantum systems is that the dissipative part of AD system corresponds to $B_{AD} = \gamma \cdot \tilde{B}$ in which $\gamma$ is positive constant, while the dissipative part of Non-Markovian quantum systems corresponds to $B_{NM}(t) = 2d(t) \cdot \tilde{B}$ in which $d(t)$ is a time-varying function. Therefore, the information lose irreversibly in AD system, while there is return information in Non-Markovian quantum systems so that the accuracy in $D$ and $F$ for Non-Markovian quantum systems should be higher than that for AD system in theory. However, the minimum $D$ of Non-Markovian system is less than that of AD system as shown in Table 2, which means the performance $D$ of Not gate preparation in AD system has higher accuracy than that in Non-Markovian quantum system. This is a contradiction that result from: 1) (24) don't use the return information effectively; 2) $B_{NM}(t)$ is time-dependent so that its effects are more difficult to overcome. As a result, the control laws will be more effective for Non-Markovian quantum systems if the return information in Non-Markovian quantum systems can be considered to design control laws.

According to the values of minimum $D$ and maximum $F$ in all systems, Not gate for PD system has highest accuracy in index $D$ in all systems, while Not gate for closed quantum system has highest accuracy in index $F$, which means the conclusion based on $D$ are different from that based on $F$. Thus indexes $D$ and $F$ are not positive correlation, and it is more comprehensive to describe the properties of systems and control laws based on two indexes. It is also important to note that Non-Markovian quantum systems takes the shortest time to the same $D$ and $F$ values in all systems, which is related with that the dissipation part of Non-Markovian quantum systems is time-dependent. The results of PD system and Non-Markovian quantum system also indicate that the system dynamics contains time-dependent dissipation part which is not always negative factors, they maybe play positive roles in some cases.

**4.4 Robustness analysis for the system Hamiltonian with uncertainty**

The control laws used in all experiments of preceding subsections are consistent with the designed values. However, the perturbations will lead to deviations in numerical values between the actual control laws and the designed control laws, as discussion in subsection 3.3. The experiments in this subsection are used to study the effects of the deviations. In the numerical simulations, we use the control laws (24) for PD and AD systems. In the cases that the perturbation $\lambda\sigma = \lambda_x\sigma_x$, $\lambda_y\sigma_y$ and $\lambda_z\sigma_z$ in (28), respectively, and $\lambda \in [-100,100]$, the values of $D$ and $F$ of the system (28) are shwon in Fig. 4, from Fig. 4 (a) one can see that the values of $D$ in the case $\lambda\sigma = \lambda_x\sigma_x$ are less than that when the perturbations are $\lambda\sigma = \lambda_y\sigma_y$, and $\lambda\sigma = \lambda_z\sigma_z$. From Fig. 4 (b) one can see that the values of $F$ when the perturbation is $\lambda\sigma = \lambda_x\sigma_x$ are more than that when the perturbations are $\lambda\sigma = \lambda_y\sigma_y$, and $\lambda\sigma = \lambda_z\sigma_z$; while the values of $D$ and $F$ when $\lambda\sigma = \lambda_y\sigma_y$ all close to that when $\lambda\sigma = \lambda_z\sigma_z$. The $F$ with $\lambda$ in the PD system is similar to that in the AD system, while the $D$ with $\lambda$ in AD system is similar to that in the PD system, so the results of those two cases do not be shown.

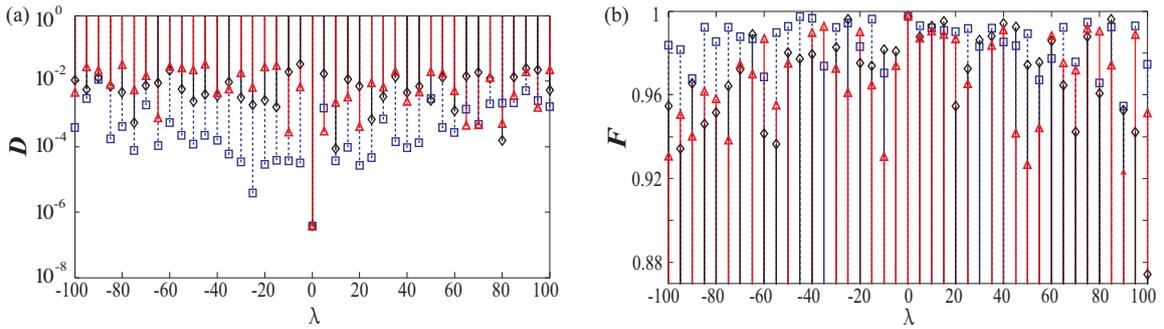

Fig. 4. $D$ and $F$ of preparing the NOT gates under the control laws (24) for PD system and AD system when deviation $\lambda \in [-100,100]$ in which '□', '△' and '◇' correspond to the cases $\lambda\sigma = \lambda_x\sigma_x$, $\lambda_y\sigma_y$ and $\lambda_z\sigma_z$, respectively; (a) $D$ with $\lambda$ in the PD system; (b) $F$ with $\lambda$ in the AD system.

One can see from Fig. 4 that the closer the values of $D$ and F to that when $\lambda=0$, the stronger the robustness, which means that in the case the perturbation is $\lambda\sigma = \lambda_x\sigma_x$ the system's robustness is stronger than that in the case $\lambda\sigma = \lambda_y\sigma_y$, and $\lambda\sigma = \lambda_z\sigma_z$; In the case $\lambda\sigma = \lambda_y\sigma_y$, the system's robustness is similar as that when $\lambda\sigma = \lambda_z\sigma_z$. The reasons are as following:

1) The action of $f_x$ is to offset the dissipation part which is from $B$. However, the value of $B$ in

the experiments is small so that the action of $f_x$ is relatively weaker position than that of $f_y$ and $f_z$ which are used to prepare operators, so the anti-interference ability of $f_x$ is stronger.

2) The actions and mathematical expressions of $f_y$ and $f_z$ are similar, so the values and changing trends are similar so that the anti-interference abilities of the two control laws are similar, and the robustness when $\lambda\sigma = \lambda_y\sigma_y$, and $\lambda\sigma = \lambda_z\sigma_z$ are also similar. Although the effect which is caused by the perturbation to the control law when $\lambda\sigma = \lambda_x\sigma_x$ is similar to the effects to control laws when $\lambda\sigma = \lambda_y\sigma_y$, and $\lambda\sigma = \lambda_z\sigma_z$, the actions of the control laws in preparing operators are different, so the robustness are also different.

As a consequence, if the directions of perturbations are known, the control laws in these directions are recommended to offset $B$; if the directions of perturbations are unknown, one can add control components which are used to offset $B$ to the three control laws at the same time as shown in (26). One can make the control laws in the perturbation directions play a greater role in offsetting $B$ through adjusting the parameters $h_{nx}$, $h_{ny}$ and $h_{nz}$ to strengthen the system robustness. In order to verify and support this conclusion, another control laws designed based on $V$ is

$$\begin{aligned} f_{yx}(t) &= -k_{yx}S(A_x) \\ f_{yy}(t) &= -\frac{S(B)}{S(A_y)}, \ k_{yx} \geq 0 \\ f_{yz}(t) &= -k_{yz}S(A_z), \ k_{yz} \geq 0 \end{aligned} \qquad (37)$$

Namely, $f_{yy}(t)$ is used to offset the dissipation while $f_{yx}(t)$ and $f_{yz}(t)$ are used to prepare operators. The $D$ and $F$ of preparing the NOT gates by (37) for PD system when $\lambda \in [-100, 100]$ are shown in Fig. 5, from which one can see that for the same $\lambda \in [-100, 100]$, the values of $D$ when the perturbation $\lambda\sigma = \lambda_y\sigma_y$ are less than that when $\lambda\sigma = \lambda_x\sigma_x$ and $\lambda\sigma = \lambda_z\sigma_z$ in Fig. (a), and the values of $F$ are more than that when $\lambda\sigma = \lambda_x\sigma_x$ and $\lambda\sigma = \lambda_z\sigma_z$ in Fig. (b), while the values of $D$ and $F$ when $\lambda\sigma = \lambda_x\sigma_x$ close to that when $\lambda\sigma = \lambda_z\sigma_z$. Using the same analysis method as Fig. 4, the results in Fig. 5 mean that the robustness when $\lambda\sigma = \lambda_y\sigma_y$ is stronger while the robustness when $\lambda\sigma = \lambda_x\sigma_x$ and $\lambda\sigma = \lambda_z\sigma_z$ is similar and weaker than that when $\lambda\sigma = \lambda_y\sigma_y$, which is consistent with the expected results.

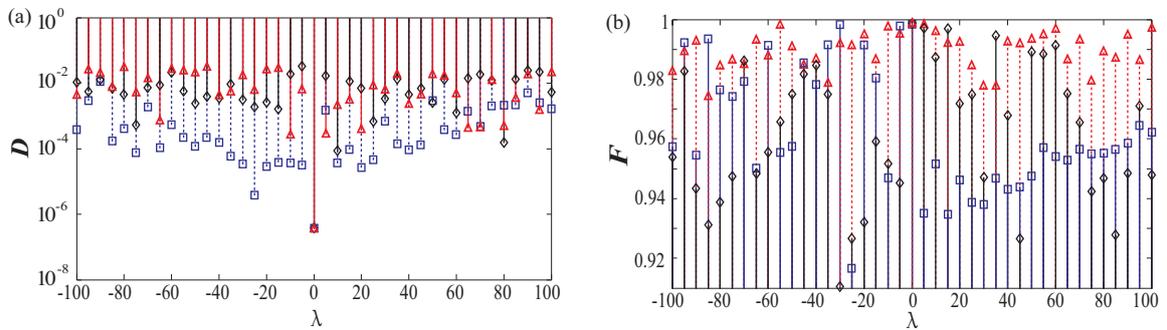

Fig. 5. $D$ and $F$ of preparing the NOT gates by the control laws (37) for PD system when $\lambda \in [-100, 100]$. '□', '▲' and '◇' correspond to the case $\lambda\sigma = \lambda_x\sigma_x$, $\lambda_y\sigma_y$ and $\lambda_z\sigma_z$, respectively, in which (a) $D$ with $\lambda$ under (37) for PD system; (b) $F$ with $\lambda$ under (37) for AD system.

## 5. Conclusions

We analyzed and prepared the operators by the control laws designed based on Lyapunov method for two level open quantum systems. A novel Lyapunov function $V$ was proposed for the operator preparation in this paper. Both the theory analysis and experiments indicate that proposed $V$ has advantages in numerical accuracy and convergence speed relative to $V_{dis}$. The control laws designed based on $V$ can prepare higher accuracy Not gates. Moreover, the control laws are appropriate for both closed quantum systems and open quantum systems, and one can choose or combine different control laws designed based on the same Lyapunov function in actual use to obtain better control performance. Besides, the robustness of system Hamiltonian with uncertainty is also investigated. The values of the control laws can be changed due to the perturbations. The experiment results for robustness show that when the value of the control law offsetting the dissipation are changed, the system has less effect based on the values of *D* and *F*, i.e. the robustness is stronger; when the values of the two control laws preparing operators are changed, the system has larger and similar effects, i.e. the robustness are weaker and similar. Based on these results, we give the suggestion for the control laws to enhance the system robustness. The proposed Lyapunov function $V$ can also be used to the quantum system state transfer, and the advantages of state transfer accuracy and speed can be expected.

## Acknowledgements

This work was supported partly by the National Key Basic Research Program under Grant No. 2011CBA00200.